\documentclass[12pt]{article} 

\usepackage{latexsym}
\usepackage{graphicx}
\usepackage{amsfonts}
\usepackage{amssymb}
\usepackage{amsmath}

\usepackage{multirow}
\usepackage{rotating}
\usepackage{lscape}

\def\be{\begin{equation}}
\def\ee{\end{equation}}
\def\ba{\begin{eqnarray}}
\def\ea{\end{eqnarray}}
\def\la{\label}

\renewcommand{\d}{{\rm d}}
\newcommand{\Tr}{\mbox{Tr\,}}

\newcommand{\beq}{\begin{equation}}
\newcommand{\eeq}[1]{\label{#1}\end{equation}}
\newcommand{\bea}{\begin{eqnarray}}
\newcommand{\eea}[1]{\label{#1}\end{eqnarray}}
\renewcommand{\Re}{\mathrm{Re}\,}
\renewcommand{\Im}{\mathrm{Im}\,}

\newcommand{\Cliff}{\operatorname{Cliff}}
\newcommand{\Cset}{\mathbb{C}}
\newcommand{\e}{{\mathrm e}}
\newcommand{\eg}{{\em e.g.}~}
\newcommand{\etc}{{\em etc}}

\newcommand{\Hdila}{H}
\newcommand{\half}{\frac{1}{2}}
\newcommand{\Hset}{\mathbb{H}}

\newcommand{\Mat}{\operatorname{Mat}}

\newcommand{\Pin}{\operatorname{Pin}}

\newcommand{\Rset}{\mathbb{R}}
\newcommand{\SL}{\operatorname{SL}}

\newcommand{\Spin}{\operatorname{Spin}}

\newcommand{\tr}{\operatorname{tr}}

\newcommand{\unit}{\mathbf{1}}
\newcommand{\Vol}{\operatorname{Vol}}
\newcommand{\Zset}{\mathbb{Z}}

\def\nn{\nonumber}

\begin{document}
\begin{flushright}\hfill{hep-th/0605273} \\
\end{flushright}

\vspace{20mm}

\begin{center}

{\bf Signature reversal invariance} \\
\vspace*{0.37truein}
M.J.~Duff\footnote{m.duff@imperial.ac.uk} and J.~Kalkkinen\footnote{j.kalkkinen@imperial.ac.uk} \\
\vspace*{0.15truein}
{\it Blackett Laboratory, Imperial College London} \\
{\it Prince Consort Road, London SW7 2AZ} \\

\end{center}

\baselineskip=10pt
\bigskip

\vspace{20pt}

\abstract{

\noindent We consider the signature reversing transformation of
the metric tensor $g_{\mu\nu} \rightarrow -g_{\mu\nu}$ induced by
the chiral transformation of the curved space gamma matrices
$\gamma_{\mu \rightarrow} \gamma \gamma_{\mu}$ in spacetimes with
signature $(S,T)$, which also induces a $(-1)^T$ spacetime
orientation reversal. We conclude: (1) It is a symmetry only for
chiral theories with $S-T= 4k$, with $k$ integer. (2) Yang-Mills
theories require dimensions $D=4k$ with T even for which even rank
antisymmentric tensor field strengths and mass terms are also
allowed. For example, $D=10$ super Yang-Mills is ruled out. (3)
Gravititational theories require dimensions $D=4k+2$ with $T$ odd,
for which the symmetry is preserved by coupling to odd rank field
strengths. In $D=10$, for example, it is a symmetry of N=1 and
Type IIB supergravity but not Type IIA. A cosmological term and
also mass terms are forbidden but non-minimal $R\phi^{2}$ coupling
is permitted. (4) Spontaneous compactification from $D=4k+2$ leads
to interesting but different symmetries in lower dimensions such
as $D=4$, so Yang-Mills terms, Kaluza-Klein masses and a
cosmological constant may then appear. As a well-known example,
IIB permits $AdS_{5} \times S^{5}$.

}

\newpage


\tableofcontents

\newpage


\section{Signature reversal}

\subsection{Bicoastal theories}
\label{Bicoastal}

An interesting parlour game is to write down the equations of motion
of
various theories and then have your friends guess what spacetime
signature you had in mind. Was it East coast conventions $(-+++)$ or
West coast $(+---)$? For example, a free massive scalar field obeying
\be
(\eta^{\mu\nu}\partial_{\mu} \partial_{\nu}+m^{2})\phi=0
\ee
is unambiguously West coast, since the East coast kinetic term has the
opposite sign. The same is true for a  Dirac spinor. If the gamma
matrices obey
\ba
\{ \gamma_\mu , \gamma_\nu \} &=& 2 \eta_{\mu\nu} \unit ~,
\la{dirac}
\ea
then on the West coast
\be
(-i\gamma^{\mu}\partial_{\mu}+m)\psi=0~,
\ee
while there is no factor of $i$ on the East. Had the fields been
massless,
on the other hand, there would have been no way to tell.

Even in the massless case, coupling to electromagnetism gives the game
away:
\ba
\partial^{\mu}F_{\mu\nu} = -e{\bar \psi}\gamma^{\mu}\psi \\
-i\gamma^{\mu}(\partial_{\mu}-ieA_{\mu})\psi = 0
\ea
is once again West coast since the right hand side of the Maxwell
equation would acquire a factor of $i$ on the East.

On the West Coast the empty space Einstein's equations with a
cosmological constant are
\be
R_{\mu\nu}-\frac{1}{2}g_{\mu\nu}R-\frac{1}{2}\Lambda g_{\mu\nu}=0
~,
\ee
but the sign of the cosmological term is opposite on the East. Had
there been no cosmological constant, on the other hand, there
would be no way to tell. In that case, moreover you could have
coupled gravity to a massless scalar field and even allowed a
non-minimal coupling
\be
(g^{\mu\nu}\nabla_{\mu} \nabla_{\nu}+\xi
R)\phi=0
\ee
and your friends would still be unable to guess the
signature. We shall refer to this class of theory as
``bicoastal''.

A natural question to ask, therefore, is whether this can be
promoted to a symmetry principle. We shall consider theories in
spacetime signature $(S,T)$ that are invariant under reversing the
sign of the metric tensor, which  may be regarded as a reversal of
signature $(S,T) \rightarrow (T,S)$. Since we prefer to transform
fields rather than constants, we work with the curved space
metric, henceforth denoted $G_{MN}(x)$, rather than the Minkowski
metric $\eta_{MN}$,  and consider the transformation
\ba
G_{MN}(x) \rightarrow -G_{MN}(x) \la{metricflip} ~.
\ea
Similarly, we shall work with the curved space gamma matrices
\ba
\Big\{ \Gamma_M(x) , \Gamma_N(x) \Big\} &=& 2 G_{MN}(x) \unit
\la{Clifford}
\ea
rather than the flat space ones.

In the presence of fermions, it is useful to implement this
reversal by the transformation
\ba
\Gamma_{M} &\rightarrow& \Gamma \Gamma_{M} ~,  \la{gammaflip}
\ea
where $\Gamma$ is the normalised chirality operator
\ba
\Gamma &\equiv & \frac{1}{\sqrt{G}} \frac{1}{D!}\varepsilon^{{M_1}
\cdots M_{D}}\Gamma_{{M_1} \cdots M_{D}} \label{chiral}
\ea
and where
\ba
\sqrt{G} & \equiv & \sqrt{ (-1)^T \det G_{MN}(x) }  ~.
\label{SqrtDetG}
\ea
Let us denote the Clifford algebra by $\Cliff(S,T)$. A change of
signature may change the Clifford algebra generated by the gamma
matrices. There are, however, special dimensions where the
Clifford algebra is isomorphic in opposite signatures. This is
necessary so that the fermion representations do not jump, and
that the fermion interactions in the action remain real. As
explained in appendix \ref{LocalFermi}, in order that the Clifford
algebra remain the same under signature reversal
\ba
\Cliff(S,T) &=& \Cliff(T,S)
\ea
we require
\ba S-T &=& 4k'
\ea
for some integer $k'$. See tables \ref{tab2}, \ref{tab3} and 
\ref{tab4}.  This rules out odd $D$, and one can express
the dimension of the spacetime in terms of an arbitrary integer
$k$ and the number of time-like directions $T$ as
\ba
D &=& 4k' + 2T ~.
\ea
By redefining $k$ we have therefore two prototypical admissible
dimensionalities:
\begin{itemize}
\item[-]
The Minkowskian type with an odd number of time-like directions
$D=4k + 2$;
\item[-]
The Euclidean type with an even number of time-like directions
$D=4k$.
\end{itemize}
In both cases,
\ba
\Gamma^2 &=& +1 \label{gammasquared} ~.
\ea
Since $\Gamma$ anticommutes with $\Gamma_{M}$ the operation
(\ref{gammaflip}) will reverse the sign of the metric as desired.
We also note that under (\ref{gammaflip})
\be
\Gamma \rightarrow (-1)^{T}\Gamma \la{Tgammaflip}
\ee

As explained in section \ref{volume} this means that
for dimensions $D=4k$ the sign of the volume form
$\Vol(M) \equiv \sqrt{G} \d^{D} x$
remains the same while for $D=4k+2$ it changes sign corresponding to
reversal of orientation. The link between the  orientation and the
signature  arises from the Clifford algebra.
In the signatures $S-T=4k'$ where Clifford algebra remains isomorphic
under change of signature, we have $D/2 \equiv T \mod 2$, and the na\"{\i}ve
transformation
\ba
\sqrt{G} ~ \d^D x & \rightarrow & (-1)^{T} ~ \sqrt{G} ~ \d^D  x
\ea
and the transformation induced by (\ref{gammaflip})
\ba
\frac{1}{D!} \varepsilon^{{M_1} \cdots M_{D}} \Gamma_{{M_1} \cdots
M_{D}}  & \rightarrow & (-1)^{T} ~ \frac{1}{D!} \varepsilon^{{M_1}
\cdots M_{D}} \Gamma_{{M_1} \cdots M_{D}}
\ea
have an identical effect. Note that the density $\sqrt{G} >0$ is always positive, and does
therefore not change sign.

The change in sign of the volume element then raises an important
question. Do we demand only that the equations of motion be
invariant, which would allow the action to change by an overall sign,
or do we insist on the stronger requirement that the action be
invariant?
Let us first consider the case of pure gravity.

\subsection{Pure gravity}

The D-dimensional gravitational action functional is
\ba
S_E &=& \frac{1}{2\kappa_{D}^{2}}\int \d^D x \sqrt{G} ~ R ~.
\ea
Under (\ref{metricflip}) the volume element transforms as
\ba
\sqrt{G} \,\d^Dx & \longrightarrow & (-1)^{D/2}\,\sqrt{G} \,\d^Dx
\la{volflip}
\ea
while the curvature scalar flips sign for all $D$
\ba
R & \longrightarrow & -R ~.
\la{Rflip}
\ea
The requirement of invariance selects out the dimensions
\ba
D &=& 4k+2~, \qquad k=0,1,2,3\ldots  \label{a4}
\ea
and forbids a bulk cosmological constant $\Lambda$ in the action
\ba
S_{\Lambda} &=& \frac{1}{2\kappa_{D}^{2}} \int \d^Dx \sqrt{G} \Lambda ~.
\ea

As we have seen, the weaker requirement, that the Einstein equations
be invariant, rules
out a cosmological constant in any dimension at the classical
level but at the quantum level there
will be $L$-loop counterterms of the form
\ba
S_c &\sim& \frac{1}{2\kappa_{D}^{2}}\int \d^D x \sqrt{G}
 \kappa_{D}^{2L} ~ R^{\frac{(D-2)L+2}{2}}
\ea
where $R^n$ is symbolic for a scalar contribution of $n$ Riemann
tensors each of dimension 2. This again requires $D=4k+2$ for
invariance. So we shall take the view that the action, and not
just the equations of motion should be invariant.

Of course, one could argue that this is giving the game away since
the requirement that the Hamiltonian be positive then forces a
specific sign for scalar and gravity kinetic terms which will flip
in the opposite signature. So, if one prefers, one could adopt the
view that a bicoastal gravitational theory is one for which the
field equations, including their quantum corrections, are
insensitive to signature flip. This would lead to the same
$D=4k+2$ requirement and, if fermions are involved, the same
$S-T=4k$ condition with $T$ odd. This point of view may be
especially compelling for theories with self-dual field strengths
such as Type IIB supergravity in $D=10$ and chiral supergravity in $D=6$
which have no Lorentz-invariant action principle.

\subsection{Comparison with other authors}
\la{comparison}

Motivated primarily by the desire to explain the (approximate)
vanishing of the
cosmological constant, several authors, Erdem
\cite{Erdem:2004yd,Erdem:2006qk},
Quiros \cite{Quiros:2004ge}, `t Hooft and Nobbenhuis
\cite{Nobbenhuis:2004wn,
'tHooft:2006rs},
Kaplan and Sundrum \cite{Kaplan:2005rr}, have recently considered the
transformation
\ba
x^{M} &\rightarrow&  i\, x^{M} \la{xflip}
\la{reversal}
\ea
with the understanding that $G_{MN}(ix)=G_{MN}(x)$. This also induces a signature reversal
\begin{eqnarray}
 \d s^2\,=\,G_{MN} ~ \d x^M\,\d x^N &\rightarrow&  -\,\d s^2 ~.
\la{sigflip}
\end{eqnarray}
Signature reversal and the cosmological constant problem was also discussed in
signature $(3,3)$ by Bonelli and Boyarsky in
\cite{Bonelli:2000tz}. For earlier work on flipping the sign of
the metric and its relation to Clifford algebras, see
\cite{Carlip:1988gw,Berg:2000ne}.

For clarity, therefore, we note the following differences and
similarities of our approach:
\begin{itemize}

\item[1)] {\it There is more to our chiral transformation (\ref{gammaflip})
than just signature reversal (\ref{sigflip}).} There is also the change
of orientation (\ref{Tgammaflip}) which depends on whether $T$ is even or
odd.  So we are excluding some theories that would be allowed by
(\ref{reversal}).

\item[2)] {\it We work only in $D=4k' + 2T$ dimensions.}
`t Hooft and Nobbenhuis \cite{Nobbenhuis:2004wn,'tHooft:2006rs}
work in four dimensions and note that the requirement that the
Einstein equations be invariant under (\ref{xflip}) rules out a
cosmological constant.  As noted by Erdem
\cite{Erdem:2004yd,Erdem:2006qk}, the stronger requirement of
invariance of the action under (\ref{xflip}) again selects out the
dimensions
\begin{equation}
D=4k+2~~~,~~~~k=0,1,2,3, \ldots
\end{equation}
In this respect we agree with Erdem
\cite{Erdem:2004yd,Erdem:2006qk} because, as noted above, quantum
corrections require this stronger condition. We differ from Erdem
in the transformation rules for matter fields, however. We also
differ from Quiros \cite{Quiros:2004ge} who compensates for the
flip of sign of the four-dimensional action by a flip in the sign
of Newton's constant.

\item[3)] {\it We consider only positive energies.}
As we have just seen, by working in $D=4k'+2T$ dimensions, we
avoid transformations that change the sign of the action and hence
avoid all discussion of ghosts and other interpretational problems
that negative energies involve. This differs from `t Hooft and
Nobbenhuis \cite{Nobbenhuis:2004wn,'tHooft:2006rs}, Kaplan and
Sundrum \cite{Kaplan:2005rr}, and the earlier work of Linde
\cite{Linde:1988ws}.

\item[4)] {\it Real coordinates transform only into real coordinates.}
By reversing the signature using a transformation on the metric
(\ref{metricflip}) and the curved space gamma matrices
(\ref{gammaflip}) we avoid the introduction of complex coordinates
or Wick rotations. (However, since in the appropriate signature the complex
transformation (\ref{reversal})
on the coordinate $x^{M}$  reproduces formally  the same transformation
rules for the volume element (\ref{volflip}) and the curvature
(\ref{Rflip}) as the chiral transformation (\ref{gammaflip}) on the curved
space gamma matrices, it is sometimes useful in providing a check on our
results.  That metric reversal  (\ref{metricflip}) yields the same transformation
rules as (\ref{reversal}) was first stated by Erdem \cite{Erdem:2006qk}, but
without the Clifford algebra result (\ref{Tgammaflip}) it is difficult to justify the
orientation reversal (\ref{volflip}) in $D=4k+2$ that changes the sign of the
volume element.)

\item[5)] {\it Real fields transform only into real fields.}
We avoid transformations which take real fields into imaginary
ones. This is important and restricts our choice of field
parameterizations. For example, the metric and curved space gamma
matrices are ``good'' variables in this respect, while the
vielbein $e_{M}^{~~A}$ for which
\ba
G_{MN} &=&  e_{M}^{~~A}e_{N}^{~~B} ~ \eta_{AB}
\ea
is a ``bad'' variable, since there is in general no real tranformation 
on the vielbeins that induces our basic transformation 
(\ref{gammaflip}). See appendix \ref{real}.  For this reason we 
avoid vielbeins and work directly with the curved-space Dirac matrices.  As shown
in section \ref{Fermions}, it is in fact possible to describe the coupling of 
fermions to gravity using just the gammas, without ever having to 
introduce
veilbeins (an interesting observation in its own right). 

Similarly, the scalar
field ${\Hdila}(x)$ whose vev is the string coupling constant
\ba
\langle {\Hdila}(x) \rangle &=& g_{s}
\ea
is a good variable whose sign we can flip \cite{Duff:2006}, while the
usual
$\Phi$ parameterization
\ba
{\Hdila} &=& \e^{\Phi}
\ea
is bad from this point of view.

\item[6)] {\it Constants do not transform.}
We do not consider transforming constants such as the Minkowski
metric $\eta_{MN}$ or flat space gamma matrices. Nor do we
transform Newton's constant \cite{Quiros:2004ge} or particle
masses \cite{Erdem:2004yd}.

\end{itemize}

In making the above comparisons, we do not wish to imply that these
alternative approaches
are without merit. We merely wish to note that, in some respects, our
approach is more
conservative.

\section{Inclusion of boson matter fields}
\label{Bose}

\subsection{Antisymmetric tensor fields}

Extension of this symmetry to include matter in $D=4k+2$ requires
that the kinetic terms should transform in the same way as the
curvature scalar, with a reversal of sign. This allows antisymmetric
tensor field strengths of odd rank but not even
\be
S_{F} \equiv \int \d^Dx \sqrt{G}
G^{M_{1}N_{1}}G^{M_{2}N_{2}}\ldots G^{M_{n}N_{n}}
F_{M_{1}M_{2}\ldots M_{n}}F_{N_{1}N_{2}\ldots N_{n}}
\ee
and also rules out all mass terms, although a non-minimal
$R\phi^{2}$ scalar coupling is allowed.  It is at this stage that
the two versions of signature reversal (\ref{metricflip}) and
(\ref{xflip}) may diverge since one is free to assign different
transformation rules to the matter fields. To keep them the same
would require covariant vectors to transform with the opposite
sign to contravariant under (\ref{reversal}).

In $D=4k$, on the other hand, signature reversal allows antisymmetric
tensor field
strengths with even rank, as well as mass terms.

\subsection{Pure Yang-Mills}

The kinetic term in pure Yang-Mills
\ba
S_{YM} &=& \frac{1}{4 g_D^2} \int \d^D x ~ \sqrt{G} ~ \Tr
|F_{2}|^{2}
\ea
contains two contractions with the background metric, and is
therefore invariant under reversal of signature. The same is true
of Chern-Simons terms as well.  Invariance then requires that the
volume form should not change sign under signature reversal.
Consequently Yang-Mills theory is invariant only in dimensions
$D=4k$. If the theory is coupled to fermions, and we require $S-T
= 4k'$, this leads to $D=4k'+2T$ so that there would have to be an
even number of time-like dimensions. This is the case in the
Euclidean four-dimensional spacetime, for instance.

Quantum corrections that involve only the field strength $F$ are
invariant under signature reversal. Corrections that involve
powers of the d'Alambertian $\Box$ need not be, however.
$L$-loop corrections that may on dimensional grounds arise in perturbation
theory are, schematically, of the form
\ba
S_{c} & \sim & \frac{1}{ g_{D}^2 } \int \d^D x ~ \sqrt{G} ~ \Tr F_{2} \Big( g_{D}^2
\Box^{(D-4)/2} \Big)^L F_{2} ~.
\ea
These terms are invariant for all $L$ under signature reversal
only when $D=4k$. This is consistent with the fact that Yang-Mills
theories are signature reversal invariant in signatures of
Euclidean type.

There are several caveats, however:
\begin{itemize}
\item[1)] Although forbidden in $D=4k+2$, Maxwell and Yang-Mills
terms can arise after compactification to lower dimensions. See,
for example, section \ref{compact}.
\item[2)] The absence in $D=4k+2$ applies in pure Yang-Mills theory
only. In the  Yang-Mills sector of Type I supergravity the kinetic
term is multiplied by a  dilaton factor ${\Hdila}(x)$ that may
also change sign to compensate for the change in sign of the
volume element \cite{Duff:2006}.
\item[3)] Yang-Mills interaction may also appear on branes, for
example D3, where the rules of signature reversal are different
from those in the bulk \cite{Duff:2006}.
\end{itemize}

\subsection{Self-duality}

A real $n$-form can be (anti-)self-dual in $D=S+T$ dimensions
provided that the Hodge star operation $\star$, which obeys
\ba
\star^2 &=& (-1)^{n(D-n) +T}
\ea
when operated on $n$-forms, is nilpotent. As the dimension $D=2n$
is even, this happens for even $n$ in signatures of Euclidean type
($T$ even), and for odd $n$ in signatures of Minkowskian type ($T$
odd). Self-dual fields arise therefore in gravity-like theories in
Minkowskian signature and in Yang-Mills like theories in Euclidean
space.

Under signature reversal the Hodge star operated on an $n$-form
$F_n$ transforms as
\ba
\star F_n &\longrightarrow & (-1)^{n+T}  \star F_n ~.
\ea
This is because the Hodge star carries information of the
orientation of the spacetime, and the volume form picks up the
sign $(-1)^T$ under signature reversal; further, the Hodge star
involves  $D-n$ lowered indices, and hence $D-n$ occurences of the
metric tensor. An (anti-)self-duality equation is therefore left
form invariant under signature reversal precisely when it is
defined.

\section{Inclusion of fermion matter fields}
\label{Fermions}

\subsection{Issues of signature}

Fermions and their interactions are sensitive to the choice of
signature in several ways. As explained above, in this paper we
restrict to such signatures  $S-T = 4k'$ and field theories which
happen to be independent of whether the actual signature is
$(S,T)$ or $(T,S)$. To appreciate the importance of making this
restriction, it is instructive to see what would happen otherwise:

First of all, fermions belong to representations of the Clifford
algebra. If the Clifford algebra changes, the minimal fermion
representations change, as discussed at length in appendix
\ref{LocalFermi}. An example of this is the three dimensional
Minkowski signature $(2,1)$,  where the Clifford algebra $\Rset(2)
\oplus \Rset(2)$ consists of pairs of real $2 \times 2$ matrices
in the mostly plus, or East coast, signature and of $2 \times 2$
complex matrices $\Cset(2)$ in the West coast signature. (A
similar statement applies in eleven dimensions as well, as both
have $S-T = 4k' +1$, and not $4k'$.)

Changing signature when  $S-T \neq 4k'$  changes the reality properties of fermion
interactions. In many cases a change of signature amounts indeed
to multiplying gamma matrices with the imaginary unit: in such
cases, if fermion interactions involve odd numbers of gamma
matrices (\eg the kinetic term), this  makes the  action imaginary
and the theory  non-unitary. In  $S-T = 4k'$ the Lagrangian is the same in both signatures.

These observations mentioned above are local in nature. The
final difference  concerns defining fermions globally. Though we
shall not discuss global issues in the present paper, we recall it
briefly here: namely, if the background spacetime is
non-orientable, the global structure of fermion fields may change
when the signature is changed \cite{Carlip:1988gw,Berg:2000ne}. In
particular, the topological obstruction to defining a global
fermion field then typically changes. As with the
differences mentioned above, this does not happen when $S-T =
4k'$.

\subsection{Clifford volume}
\la{volume}

In section \ref{Bicoastal} we argued that the transformation
properties of the volume element should be consistent with the
representation of the volume form in terms of elements of the
Clifford algebra. We shall here show in more detail in what sense
the chirality operator $\Gamma$ and the volume element $\sqrt{G}
\d^{D} x$ can be identified.

Arbitrary linear combinations of products of gamma-matrices
generate the full Clifford algebra $\Cliff(TM)$. At a given point
$x \in M$ the Clifford algebra is a vector space with the basis
\ba
\unit, \Gamma_{M}, \Gamma_{MN}, \ldots, \Gamma_{M_{1} \cdots
M_{D-1}}, \Gamma ~, \label{basis}
\ea
where $\Gamma$ was defined in (\ref{chiral}). In even dimensions,
only $\unit$ has a non-zero trace. A typical element $\Sigma$ at
degree $n$ of the Clifford algebra $\Cliff(TM)$ can then be
expanded locally as
\ba
\Sigma &=& \frac{1}{n!} \Sigma_{M_1 \cdots M_n} \Gamma^{M_1 \cdots
M_n} ~.
\ea
As a vector bundle on $M$, the Clifford algebra is the same as the
space of differential forms
\ba
\Cliff(TM) &\simeq& \Omega^*(M) ~. \label{isomo}
\ea
The multiplicative structure is different: Clifford product
`$\cdot$' in the former, wedge product `$\wedge$' in the latter.

As we have the metric $G_{MN}$ at our disposal, we may define the
Hodge dual of a differential form as well as of an element of the
Clifford algebra
\ba
\star \Sigma &=& \frac{(-1)^{T}}{(D-n)! n! ~ \sqrt{G}}
 ~ {\varepsilon^{M_1 \cdots M_n}}_{N_{n+1} \cdots N_D
} ~ \Sigma_{M_1 \cdots M_n}\Gamma^{N_{n+1} \cdots N_D} ~.
\label{Hodge}
\ea
The definition of $\sqrt{G}$ was given in (\ref{SqrtDetG});  the
orientation implied  in $\d^D x$ is
\ba
\d x^{M_{1}} \wedge \cdots \wedge \d x^{M_{D}} &=&
\epsilon^{M_{1}\cdots M_{D}} \d^D x ~.
\ea
This determines the standard volume element
\ba
\Vol(M) &=& \star 1 \equiv \sqrt{G} \ \d^D x \\
 &=& (-1)^{T} ~ \Gamma \label{vG}
\ea
that can be expressed as a differential form as well as an element
of the Clifford algebra.

\subsection{Kinetic terms}
\la{kinetic}

Defining kinetic terms and interaction potentials for fermions
requires one piece of additional structure, a spin (or pin)
invariant inner product
\ba
\overline\psi \chi &\equiv& \psi^{\dagger} C \chi ~,
\ea
where $C$ is a constant matrix. As we are working in $S-T = 4k'$ a
multiple of four, the dagger ${}^{\dagger}$ stands for transpose
for even $k'$ and for quaternionic conjugation when $k'$ is
odd.\footnote{In terms of imaginary Pauli matrices, this is of
course just Hermitean conjugation.} The inner product must be
invariant in order that the Lagrangean not break Lorentz symmetry,
and in order that the Leibnitz rule for a spin connection hold.

There are generally two such invariant inner products: given one
$C$ the other candidate is proportional to $\Gamma C$. A
convenient way to distinguish these inner products \cite{RH} is to
keep track of the Hermitean conjugate of a rank $n$ Clifford
matrix $\Sigma$
\ba
(\Sigma \psi)^{\dagger} \check{C} \chi &=&
(-1)^{\half n(n-1)} ~ \psi^{\dagger} \check{C} \Sigma \chi \\
(\Sigma \psi)^{\dagger} \hat{C} \chi &=& (-1)^{\half n(n+1)} ~
\psi^{\dagger} \hat{C} \Sigma \chi ~.
\ea

We must choose such an inner product that the fermion kinetic
terms
\ba
\overline{\lambda} \Gamma^M D_M \lambda \quad \text{ and }
\quad \overline{\psi_M} \Gamma^{MNP} D_N \psi_P
\ea
for the dilatino and the gravitino are nontrivial. Keeping in mind
that fermion fields are Grassmann odd, this leads to a symmetry
requirement for the inner product
\begin{itemize}
\item If $C=\check{C}$ the inner product must be symmetric
$\check{C}^{T}=\check{C}$;
\item If $C=\hat{C}$ the inner product must be anti-symmetric
$\hat{C}^{T}=-\hat{C}$.
\end{itemize}
This is not always possible. We shall here discuss the
implementation of this separately in $D=4k$ and $D=4k+2$
dimensions. The results have been summarised in table
\ref{projsum}.

\renewcommand{\arraystretch}{1.7}
\begin{table}
\begin{center}
\begin{tabular}{|c|cccc|cc|}
\hline \multirow{2}{5pt}{${\cal P}$} &   \multirow{2}{30pt}{$n=0
$} &  \multirow{2}{30pt}{$n=1$} &  \multirow{2}{30pt}{$n=2$} &
 \multirow{2}{30pt}{$n=3$} & \multicolumn{2}{c|}{$n=2k+1$}
 \\
 &&&&&   $k$ even & $k$ odd   \\
 \hline
$D=4k+2$ & $0$ & ${\cal P}_- $ & $\unit$& ${\cal P}_+ $ & ASD & SD \\
$D=4k$ & $\unit$ & ${\cal P}_+$ & $0$ & ${\cal P}_-$ & $\unit$ &  $0$ \\
 \hline
\end{tabular}
\end{center}
\caption{Projections ${\cal P}$ appearing in invariant Yukawa
couplings $\bar\psi F_n {\cal P}\chi$, where (A)SD refers to the
projection to the (anti-)self-dual component, and where in the
middle column $n$ is the rank of the tensor $F_n$ modulo four.
\label{projsum}}
\end{table}

\subsubsection{Minkowskian type $D=4k+2$}

In dimensions $D=4k+2$ we have an odd number of time-like
directions. The typical example is the Minkowski signature: it
turns out that in that signature  both of the inner products
$\check{C}$ and $\hat{C}$ have the symmetry property we required
in section \ref{kinetic}. To be  concrete, these Minkowski
signatures are $(S,T) = (1,1),(5,1)$ and $(9,1)$ in dimensions
less or equal to 14.

Apart form the Minkowski case, the middle-dimensional signature
with $S=T$ is interesting. The symmetry property for inner
products $\check{C}$ and $\hat{C}$ is satisfied then only if the
number of time-like directions is $T=1$ modulo  $4$. This means
that $(S,T) =  (1,1)$ and $(5,5)$ have a nontrivial kinetic term,
but $(S,T) = (3,3)$ and $(7,3)$ do not. This does not prohibit us
from writing down equations of motion for them, though.

These turn out to be all the admissible signatures with $D =4k+2
\leq 12$. We are now ready to discuss the behaviour of kinetic
terms under change of signature. In the above signatures, the
invariant fermion kinetic terms are
\ba
&& \int \d^D x   \sqrt{G} ~ \Big( \overline{{\cal P}_-\lambda}
\Gamma^M D_M
\lambda \Big) \label{invD} \\
&&  \int \d^D x  \sqrt{G} ~ \Big( \overline{{\cal P}_+\psi_M}
\Gamma^{MNP} D_N \psi_P \Big) ~. \label{invG}
\ea
The gravitino and the dilatino must therefore be Weyl fermions of
opposite chirality. This is indeed the case in Type IIB and Type I
supergravities in $D=10$, as well as in chiral
supergravity in $D=6$. Type IIA, for instance, does not have this
property.

\subsubsection{Euclidean type $D=4k$}

When $D=4k$ we have an even number of time-like directions. The
prime example of this is the Euclidean signature with $T=0$. Then
only $\check{C}$ produces non-trivial kinetic terms for fermions,
and the admissible signatures are $(S,T) = (4,0), (8,0)$ and
$(12,0)$.

In the middle dimensions we have signatures $(S,T) = (2,2)$ and
$(6,6)$ with $\hat{C}$, and $(S,T) = (4,4)$ with  $\check{C}$. The
other admissible signatures in $D =4k \leq 12$ that have not been
mentioned above are $(S,T) = (6,2), (10,2)$ and $(8,4)$.

The  invariant fermionic kinetic terms are the same as in
(\ref{invD}) and (\ref{invG}). This is because the volume element
$\sqrt{G}  ~ \d^D x $ picks up the opposite sign in the change of
signature to compensate for the fact that the chirality operator
$\Gamma$ picks up the opposite sign under conjugation.

\subsection{Yukawa coupling to tensors}
\la{Yukawa}

The isomorphism (\ref{isomo}) between differential forms and the
Clifford algebra enables one to write actions for form fields in
terms of the corresponding Clifford matrices. This relies on the
fact that  (\ref{basis}) is an orthogonal basis. For the rank-$n$
field $F_n$ we can therefore write the action
\ba
{\int  \frac{1}{g^2} F_n \wedge \star F_n + \theta F_n \wedge F_n}
& \sim & \int \sqrt{G} \ \d^D x \ \Tr \Big( \frac{1}{g^2} \unit +
\theta\Gamma \Big) \cdot  {\cal F}_n \cdot {\cal F}_n   ~,
\label{modelFF}
\ea
where $g^2$ and $\theta$ are arbitrary couplings, and `$\cdot$'
stands for Clifford multiplication. To emphasise the r\^ole played
by the gamma matrices, we write
\ba
F_n & \equiv & \frac{1}{n!} F_{M_1 \cdots M_n} \d x^{M_1} \wedge
\cdots \wedge \d x^{M_n} \\
{\cal F}_n & \equiv & \frac{1}{n!} F_{M_1 \cdots M_n} \Gamma^{M_1 \cdots M_n}
~.
\ea
The topological term proportional to $\theta$ vanishes
consistently on both sides when $D \neq 2n$ or $n$ is odd.

This allows one to revisit the symmetry properties of the form
field action under signature reversal. For this we note first that
on even tensors $F_n$ with rank $n$ a change of signature induces
a change of sign
\ba
{\cal F}_n & \longrightarrow & (-1)^{n/2} {\cal F}_n ~.
\ea
This means that for even $n$ the invariance properties of the
action follow purely from the properties of $\Gamma$ and $\sqrt{G}
 ~ \d^D x $: the kinetic term is invariant only when $\sqrt{G} ~ \d^D x $ is, whereas
the topological term is always invariant.

If the rank $n$ is odd and the dimension $D$ even, we get
\ba
{\cal F}_n & \longrightarrow & - \star {\cal F}_n ~.
\ea
Using the standard results
\ba
\Gamma ~ {\cal F}_n &=& (-1)^{\half n(n+1) + n D} (\star {\cal F}_n ) \\
\star^2 {\cal F}_n &=& (-1)^{n(D-n) + T } {\cal F}_n
\ea
one can show that the action transforms to
\ba
\int \sqrt{G} \ \d^D x \Tr \Big( (-1)^{T+n} \frac{1}{g^2} +
\theta\Gamma \Big) \cdot {\cal F}_n \cdot {\cal F}_n   ~.
\ea
This reproduces the result of section \ref{Bose} that followed
simply from counting the number of times one had to use the metric
to raise indices in the kinetic term: we find that odd-rank
theories when $T$ is odd, such as in Minkowski signature, and even
rank theories when $T$ is even.

Yukawa couplings are of the form
\ba
 \int \d^D x \sqrt{G} ~ F_{M_1 \cdots M_n} ~ \overline{\psi}\Gamma^{M_1 \cdots
M_n} \chi. \label{modelxxF}
\ea
One of the reasons why we had to restrict to dimensions $S-T =
4k'$ was that these interactions should remain unitary under
change of signature, and this can be guaranteed only where the
Clifford algebra is isomorphic in opposite signatures. One can
show that the Yukawa couplings transform to
\ba
(-1)^{\half n(n-1)+T}  ~  \int \d^D x \sqrt{G} ~ F_{M_1 \cdots
M_n} ~ \overline{\psi}\Gamma^{M_1 \cdots M_n} (\Gamma)^{n} \chi ~
~.
\ea
The factor $(-1)^{T}$ comes from (\ref{Tgammaflip}). Note that
\ba
\Gamma^M \longrightarrow - \Gamma\Gamma^M
\ea
with a contravariant index.

\subsubsection{Minkowskian type $D=4k+2$}

Yukawa type interactions are not admissible for $n = 0 \mod 4$
when the number of time-like directions is odd, as is the case
here. This excludes mass terms, for instance. The other even-rank
interactions with $n = 2 \mod 4 $ are automatically invariant in
this dimensionality.

Changing signature in tensors that have odd rank introduces a
chirality operator in them as we saw already in discussing kinetic
terms in section \ref{kinetic}. Restricting to the part of the
interaction term invariant under signature reversal leads
therefore to chirality projections. When the rank of the tensor is
$n = 1 \mod 4$ even terms are of the form
\ba
 \int \d^D x  \sqrt{G} ~ F_{M_1 \cdots M_n} ~ \overline{\psi}\Gamma^{M_1 \cdots
M_n} {\cal P}_{-} \chi . \label{aliveM}
\ea
Note that, as $n$ is odd, only the negative chirality component of
$\psi$ couples here to the tensor field. For $n = 3 \mod 4$ the
chirality projection is opposite
\ba
 \int \d^D x \sqrt{G} ~ F_{M_1 \cdots M_n} ~ \overline{\psi}\Gamma^{M_1 \cdots
M_n} {\cal P}_{+} \chi ~,
\ea
and only the positive chirality components of $\psi$ and $\chi$
are concerned.

If the rank of the tensor is precisely half of the dimension of
the spacetime $n=2k+1$, the presence of these projection operators
means that only the self-dual, or the anti-self-dual, part of the
tensor field couples to fermions. Which one it should be, depends
on whether $k$ is even or odd. If $k$ should be odd, such as is
the case for instance in the six-dimensional Minkowski signature,
it is the anti-self-dual field
\ba
{\cal F}_n &=& - \star {\cal F}_n \\
&=& (-1)^{k} ~ \Gamma \cdot {\cal F}_n
\ea
that couples to the (spin-half) fermions, consistent with the case $n = 3 \mod
4$ above. (For even $k$, $n = 1 \mod 4$ and fermions couple to the
self-dual part.) This anti-self-dual field $F_n$ is indeed odd
under the change of signature as a Clifford matrix, so that in
$D=4k+2$ the Yukawa coupling is invariant. Note that in this case
the kinetic term (\ref{modelFF}) vanishes, however, and there is
no covariant action principle.

\subsubsection{Euclidean type $D=4k$}

In this dimensionality the number of time-like directions is even
and interactions with $n = 0 \mod 4$  are symmetric under
signature reversal. This means in particular that mass terms are
invariant under it. The other even-rank interactions with $n = 2
\mod 4 $ are not invariant, however. The chirality projection in
an invariant interaction term with  $n = 1 \mod 4$ is
\ba
 \int \d^D x \sqrt{G} ~ F_{M_1 \cdots M_n} ~ \overline{\psi}\Gamma^{M_1 \cdots
M_n} {\cal P}_{+} \chi ~. 
\ea
In $n = 3 \mod 4$ the chirality projection is again opposite
\ba
 \int \d^D x \sqrt{G} ~ F_{M_1 \cdots M_n} ~ \overline{\psi}\Gamma^{M_1 \cdots
M_n} {\cal P}_{-} \chi .
\ea

The fermions appearing in these interactions must have opposite
chiralities: for instance in the $n = 3 \mod 4$ case $\chi$ has
negative and $\psi$ has positive chirality. As the invariant
kinetic terms for all spin-half fields involve the same negative
chirality projection, it follows that these odd-rank Yukawa
couplings are trivial as such in $D=4k$. However, when one of the
fermions is a Rarita-Schwinger field $\psi_M$, the interaction is
nontrivial: in $n = 1 \mod 4$
\ba
 \int \d^D x \sqrt{G} ~ F_{M_0 M_1 \cdots M_n} ~ \overline{\psi}^{M_0}\Gamma^{M_1
\cdots M_n} {\cal P}_{-} \chi ~,
\ea
as $\Gamma \psi_M = + \psi_M$ and $\Gamma \chi = - \chi$. (Now
$\chi$ has negative chirality as we introduced a new occurrence of
the metric in $\psi^M = G^{MN} \psi_N$.) In $n = 3 \mod 4$ the
r\^oles are interchanged, and we have $\psi$ and $\chi_M$.

In $D=4k$ the middle-dimensional form field with $n=2k$ picks up
the sign $(-1)^k$ under signature reversal. This merely reproduces
the result that $n = 0 \mod 4$ couplings are invariant, and  $n =
2 \mod 4$ are not. In particular, where these couplings are
consistent in $D=8l$ and $n=4l$, both the self-dual and the
anti-self-dual part of the tensor couple to fermions.

This is to be contrasted with the fact that only the
(anti-)self-dual parts of the middle-dimensional forms in $D=4k+2$
coupled to fermions. No such halving of degrees of freedom arises
in $D=4k$. This is related to the fact that in $D=4k+2$ the
fermions in a Yukawa coupling had the same chirality, and one
could in fact set $\psi=\chi$; in $D=4k$ the two fermions must
have opposite chiralities, and will therefore have to be
independent, or else the Yukawa coupling is trivial.

\subsection{Fermions without vielbeins}
\la{vielbein}

So far, we have managed to discuss both kinetic and interaction terms for
fermions without having to mention vielbeins, but working only with
the curved space gamma matrices.  Since, as explained in
Section \ref{comparison}, we wish to avoid the introduction of vielbeins
altogther, we show in this section how, in second order formalism, the covariant
derivative
is related to these curved space gammas.

The spin covariant derivative is defined in terms
of how it acts on fermions $\psi$:
\ba
D_M \psi &\equiv& \Big( \partial_M +\Omega_M \Big) \psi ~.
\ea
The gamma matrices
$\Gamma_M$ and the spin-connection $D_M$ are required to be
compatible in the sense that
\ba
\partial_M \Gamma_N - \Gamma^K_{MN} \Gamma_K  +  [\Omega_M, ~
\Gamma_N] &=& 0 \label{compa} ~,
\ea
where $\Gamma^K_{MN}$ is the Christoffel symbol. For this
equation to have solutions we notice that $\partial_M
\Gamma_N$ should be expressible in terms of a linear combination
of the original gamma matrices, for instance $\partial_M \Gamma_N
= a^{K}_{MN} \Gamma_{K}$, which indeed follows from the Clifford
algebra (\ref{Clifford}).
One can then verify that the vielbein-independent expression
\ba
\Omega_M &=& \frac{1}{4} \Big(\Gamma_{N}  \partial_M \Gamma^{N}+
\partial_M \log \sqrt{G} ~ \unit + \Gamma^{KL}
\partial_{[K} G_{L]M} \Big) \label{defi}
\ea
reproduces the usual torsionless relation between spin connections
and vielbeins when expressed in a vielbein basis.

Now that we have the curved space gamma matrices $\Gamma_{M}$ and the 
corresponding compatible spin connection $\Omega_{M}$ at our disposal, 
it is always possible to write a Lagrangian involving couplings to spinors 
in a form that does not involve explicit use of vielbeins, as long as all 
the tensor fields in the theory are written in terms of world indices 
rather than tangent space.

As a concrete example, let us consider $D=10,N=1$ supergravity in 
Einstein frame  
\ba
S_{N=1} &=&  \frac{1}{2\kappa_{10}^{2}} \int \d^{10}x ~ \sqrt{G} \Big[
  R 
  + \bar{\psi}_{M} \Gamma^{MNK}D_{N}{\psi}_{K} 
  + \frac{3}{2} H^{-2} H_{MNK}^{2} 
  + \bar{\lambda} \Gamma^{M}D_{M}\lambda \nonumber \\ 
&&   
  + \half \Big(\partial_{M} \ln H^{-2}\Big)^{2} - \frac{1}{\sqrt{2}} 
  \bar{\psi}_{M}  \Big(\Gamma^{N}\partial_{N} \ln H^{-2}\Big)  
  \Gamma^{M} \lambda  \\ 
&&  
  + \frac{\sqrt{2}}{8} H^{-1} H_{MNK} \Big( \bar{\psi}_{L} \Gamma^{MNKLR} 
  \psi_{R} - 6  \bar{\psi}^{M} \Gamma^{N} \psi^{K} + 
  \sqrt{2} \bar{\psi}_{L}\Gamma^{MNK}\Gamma^{L}\lambda \Big) \Big] ~. 
  \nonumber
\ea 
We have identified $\phi^{3/4} \equiv H$ in the notation of 
\cite{Bergshoeff:1981um}. 

The supersymmetry transformation rule for the graviton can be expressed as 
\ba
\delta G_{MN} &=& \overline{\varepsilon} \Gamma_{(M} \Psi_{N)} ~.
\ea 
The curved gamma matrices transform under supersymmetry as well: 
\ba
\delta \Gamma_{M} &=& \Big( \half \overline{\varepsilon} \Gamma^{N} \Psi_{M} 
\Big) ~ \Gamma_{N} ~.
\ea
Note that the rule is of the same form as  $\partial_M \Gamma_N
= a^{K}_{MN} \Gamma_{K}$ and that it does not involve explicit vielbeins. 
The rest of the supersymmetry transformation rules are \cite{Bergshoeff:1981um}  
\ba
\delta  H &=& - \frac{1}{2\sqrt{2}} \bar\eta\lambda H \\
\delta  C_{MN} &=& \frac{1}{2\sqrt{2}} H \Big( \bar\eta \Gamma_{M} \psi_{N} 
-\bar\eta \Gamma_{N} \psi_{M} - 
\frac{1}{\sqrt{2}} \bar\eta \Gamma_{MN} \lambda \Big) \\
\delta  \lambda  &=& -
\frac{1}{\sqrt{2}} \Big( \Gamma^{N}\partial_{N} \ln H \Big) \eta + 
\frac{1}{8} H^{-1} \Gamma^{MNK} \eta \hat{H}_{MNK}\\
\delta  \psi_{M} &=&  \hat{D}_{M} \eta + \frac{\sqrt{2}}{32} H^{-1} 
\Big( {\Gamma_{M}}^{NKL} - 9 \delta_{M}^{N}\Gamma^{KL} \Big) \eta 
\hat{H}_{NKL}
  \\
&& 
- \frac{1}{512} \Big( {\Gamma_{M}}^{NKL} - 5 \delta_{M}^{N}\Gamma^{KL} \Big) 
\eta \bar\lambda \Gamma_{NKL} \lambda + \frac{\sqrt{2}}{96}  \Big[  
\Big( \bar\psi_{M} \Gamma_{NK} \lambda \Big) \Gamma^{NK} \eta \nonumber 
 \\
&& 
+ \Big(\bar\lambda \Gamma_{NK} \eta \Big) \Gamma^{NK} \psi_{M} +
2 \Big(\bar\psi_{M}   \lambda \Big)  \eta -2 \Big(\bar\lambda \eta \Big) \psi_{M}
+ 4\Big(\bar\psi_{M} \Gamma_{N} \eta \Big) \Gamma^{N} \lambda  
\Big]  \nonumber ~. 
\ea
Since supersymmetry is sensistive to the number of degrees of freedom 
this indicates that $\Gamma_{M}(x)$ correctly propagates the same number of 
physical degrees of freedom as the metric $G_{MN}(x)$. Although 
we do not attempt it here, it would be interesting to generalize this 
approach to superspace.

\section{Example: D=10 supergravities}

\subsection{N=1 supergravity}
\la{1}

The bosonic part of the N=1 supergravity action is given by
\cite{Polchinski:1998rr}

\ba
S_{NS} &=& \frac{1}{2\kappa_{10}^{2}}\int \d^{10}x \sqrt{G}
{\Hdila}^{-2}
\left (R+4{\Hdila}^{-2}(\partial {\Hdila})^{2}-\frac{1}{12}|H_{3}|^{2} \right) \\
\ea
where
\ba
H_{3} &=& \d B_{2} ~.
\ea
This is invariant under signature flip since the change in sign of
the volume element is compensated by a change in sign of the
Einstein term and the scalar and 2-form kinetic terms.  It remains
invariant when the fermions are included according to the rules of
sections \ref{kinetic}, \ref{Yukawa} and \ref{vielbein}. The 
supersymmetry transformation rules given in \ref{vielbein} are also 
invariant.

Let us now compare this with heterotic supergravity whose action
is given by the inclusion of a Yang-Mills term
\cite{Polchinski:1998rr}
\ba
S_{het} &=& \frac{1}{2\kappa_{10}^{2}} \int \d^{10}x  \sqrt{G}
{\Hdila}^{-2} \left (R+4{\Hdila}^{-2}(\partial
{\Hdila})^{2}-\frac{1}{12}|\tilde H_{3}|^{2}
-\frac{\kappa_{10}^{2}}{g_{10}^2} ~ \Tr|F_{2}|^{2}\right)
\nonumber \\
\ea
where
\ba
\tilde H_{3} &=& \d
B_{2}-\frac{\kappa_{10}^{2}}{g_{10}^{2}}\omega_{3} ~.
\ea
Since the Yang-Mills term does not change sign, the action is no
longer invariant.

\subsection{Type IIB supergravity versus Type IIA}

The bosonic part of the Type IIB supergravity action is given by
\cite{Polchinski:1998rr}
\ba
S_{IIB} &=& S_{NS}+S_{R}+S_{CS}
\ea
\ba
S_{NS} &=& \frac{1}{2\kappa_{10}^{2}}\int \d^{10}x \sqrt{G}
{\Hdila}^{-2}
\left (R+4{\Hdila}^{-2}(\partial {\Hdila})^{2}-\frac{1}{12}|H_{3}|^{2} \right) \\
S_{R} &=& -\frac{1}{4\kappa_{10}^{2}}\int \d^{10}x  \sqrt{G} \left
(|F_{1}|^{2}+|\tilde F_{3}|^{2}+\frac{1}{2}|\tilde F_{5}|^{2}
\right ) \\
S_{CS} &=& -\frac{1}{4\kappa_{10}^{2}}\int C_{4}\wedge H_{3}\wedge
F_{3} ~,
\ea
where
\ba
F_{p+1} &=& \d C_{p} \\
\tilde F_{3} &=&F_{3}-C_{0}\wedge H_{3} \\
\tilde F_{5} &=& F_{5}-\frac{1}{2}C_{2}\wedge H_{3}+\frac{1}{2}
B_{2} \wedge F_{3}
\ea
and where we must impose the extra self-duality constraint
\ba
\star \tilde F_{5} &=& \tilde F_{5} ~.
\ea
Since both $S_{NS}$ and $S_{R}$ contain field strengths only of
odd rank, it is invariant under signature reversal.

To exhibit the $SL(2,R)$ symmetry, it is more convenient to change
to Einstein frame and define
\ba
G^{E}{}_{MN} &=& {\Hdila}^{-1/2}G_{MN} \\
\tau &=& C_{0}+i{\Hdila}^{-1} \\
{\cal M}_{ij} &=& \frac{1}{\Im  \tau}
\begin{pmatrix}
|\tau|^{2}&-\Re \tau \\
         - \Re \tau &1
\end{pmatrix} \\
F^{i}_{3} &=& \left(
\begin{array}{c}
H_{3}\\
F_{3}
\end{array}
\right) ~.
\ea
Then
\ba
S_{IIB} &=& \frac{1}{2\kappa_{10}^{2}}\int \d^{10}x
 \sqrt{G} \left ( R^{E}-\frac{\partial {\bar \tau} \partial
\tau}{2(\Im\tau)^{2}} -\frac{1}{2}{\cal
M}_{ij}F^{i}_{3}.F^{j}_{3}-\frac{1}{4}|\tilde F_{5}|^{2}\right )
\nonumber \\
&&  -\frac{1}{8\kappa_{10}^{2}}\epsilon_{ij}\int C_{4} \wedge
F^{j}_{3} \wedge  F^{j}_{3} ~,
\ea

In Einstein frame the fermion interactions are included via the
complex supercovariant quantities \cite{Schwarz:1983qr}
\ba
\hat{F}_{M} &=& {F}_{M} - \kappa^{2} ~
\overline{\psi^{*}_{M}}\lambda
\\
\hat{\Omega}_{MNP} &=& {\Omega}_{MNP} - \frac{3}{2} \kappa^{2} ~
\Big(
 \overline{\psi}_{[M} \Gamma_{N} (i\psi_{P]}) -
\overline{\psi^{*}}_{[M} \Gamma_{N} (i\psi_{P]}^{*}) \Big)\\
\hat{F}_{MNP} &=& {F}_{MNP} - 3 \kappa ~ \overline{\psi}_{[M}
\Gamma_{NP]} \lambda + 6 \kappa ~ \overline{(i\psi^{*}_{[M})}
\Gamma_{N} \psi_{P]}
 \\
\hat{F}_{MNPQR} &=& {F}_{MNPQR} - 5 \kappa ~ \overline{\psi}_{[M}
\Gamma_{NPQ} \psi_{R]} - \frac{1}{16} \kappa ~ \overline{\lambda}
\Gamma_{MNPQR} \lambda ~.
\ea
From sections \ref{kinetic} and \ref{Yukawa}, we see that the form
of these supercovariant quantities as well as the supersymmetry
transformations themselves are left invariant under change of
signature provided $F_{MNP}$ is odd under signature reversal, and
both $\psi_M$ and $\lambda$ are invariant. The fact that the
3-forms are odd under signature reversal is a consequence of
covariance under T-duality, and of supersymmetry. With these
conventions, we conclude that Type IIB supergravity is invariant
under change of signature.

We contrast this metric reversal invariance with Type IIA
supergravity, whose action is given by \cite{Polchinski:1998rr}
\be S_{IIA}= S_{NS}+S_{R}+S_{CS} \ee
where
\ba
S_{NS} &=& \frac{1}{2\kappa_{10}^{2}}\int \d^{10}x  \sqrt{G}
{\Hdila}^{-2} \left (R+4{\Hdila}^{-2}(\partial
{\Hdila})^{2}-\frac{1}{12}|H_{3}|^{2}
\right) \\
S_{R} &=& -\frac{1}{4\kappa_{10}^{2}}\int \d^{10}x  \sqrt{G} \left
(|F_{2}|^{2}+|\tilde F_{4}|^{2} \right ) \\
S_{CS} &=& -\frac{1}{4\kappa_{10}^{2}}\int B_{2}\wedge F_{4}\wedge
F_{4}~,
\ea
and where
\ba
\tilde F_{4} &=& \d C_{3}-C_{1} \wedge H_{3} ~. 
\ea
Since $S_{R}$ contains RR field strengths of even rank, it is not
invariant under signature reversal.

\section{Spontaneous compactification}
\label{compact}

\subsection{Example of dimensional reduction from $D=6$ to $D=4$}

Since signature reversal for gravity is allowed in $D=6$ but not
in $D=4$, it is of interest to examine the effects of dimensional
reduction. To see this, we look at $N=1$
supergravity. The equations of motion of both the gravity multiplet $(G_{MN},
{\cal P}_{+} \psi_{M}, B_{MN}^{+})$ and the tensor multiplet $( B_{MN}^{-},
{\cal P}_{-} \chi, \varphi)$ are separately signature reversal
invariant. As a simple example we consider the two combined.

Denoting the $D=6$ spacetime indices by $(M,N=0,\ldots,5)$, the bosonic part of
the action takes the form
\ba
S_{N=1} &=& \frac{1}{2\kappa_{6}^2}\int \d^6x
\sqrt{G}{\Hdila}^{-2}\Big[ R_G+4{\Hdila}^{-2}G^{MN}\partial_M
{\Hdila}\partial_N {\Hdila} \nonumber \\
&& \qquad\qquad\qquad 
-\frac{1}{12}G^{MQ}G^{NR}G^{PS}H_{MNP}H_{QRS}\Big]\ . \la{H}
\ea
Note that there is no self-duality condition on $H_{MNP}$.
The metric $G_{MN}$ is related to the canonical Einstein metric
$G^E{}_{MN}$ by \be G_{MN}={\Hdila}G^E{}_{MN}\ . \la{metricE} \ee
The combination of the six-dimensional $N=1$ supergravity and
tensor multiplets reduce to give the $D=4$, $N=2$ graviton
multiplet with helicities $(\pm2,2(\pm{3/2}),\pm1)$ and three
vector multiplets with helicities $(\pm1,2(\pm{1/2}),2(0))$.  In
order to make this explicit, we use a standard decomposition of
the six-dimensional metric
\begin{equation}
G_{MN}=\begin{pmatrix}
g_{\mu\nu}+A_\mu^mA_\nu^nG_{mn}&A_\mu^mG_{mn} \\
A_\nu^nG_{mn}&G_{mn}
\end{pmatrix} \ , \label{metricred}
\end{equation}
where the spacetime indices are ${\mu,\nu}=0,1,2,3$ and the
internal indices are $m,n=1,2$.  The remaining two vectors arise
from the reduced $B$ field
\begin{equation}
B_{MN}=\begin{pmatrix}B_{\mu\nu}+\half(A_\mu^mB_{m\nu}+B_{\mu
n}A_\nu^n)& B_{\mu n}+A_\mu^mB_{mn} \\
B_{m\nu}+B_{mn}A_\nu^n&B_{mn}
\end{pmatrix}
\ .
\end{equation}
Four of the six resulting scalars are moduli of the 2-torus.  We
parametrize the internal metric and 2-form as
\be
G_{mn}=CB^{-1}\left(
\begin{array}{cc}
C^{-2}+c^2&-c\\
-c&1
\end{array}\right)\ ,
\la{scalars} \ee and \be B_{mn}=b\,\epsilon_{mn}\ . \ee The
four-dimensional metric, given by $g_{\mu\nu}$, is related to the
four-dimensional canonical Einstein metric, $g^E_{\mu\nu}$, by \be
g_{\mu\nu}=Ag^E{}_{\mu\nu} ~,  \la{4DEinstein} \ee where $A$ is
the four-dimensional shifted dilaton: \be
A^{-1}=\int \d x^{5}\d x^{6}{\Hdila}^{-2}\sqrt{\det G_{mn}}={\Hdila}^{-2}B^{-1}\ . \ee
Thus the remaining two scalars are the dilaton $A$ and axion $a$,
where the axion field $a$ is defined by
\be \epsilon^{\mu\nu\rho\sigma}\partial_{\sigma}a=
\sqrt{g}A^{-1}g^{\mu\sigma}g^{\nu\lambda}g^{\rho\tau}
H_{\sigma\lambda\tau}\ , \label{eq:Haxion} \ee
and where
\ba
H_{\sigma\lambda\tau}&=&3
\Big(\partial_{[\sigma}B_{\lambda\tau]}+\frac{1}{2} A_{[\sigma}^m
F_{\lambda\tau] m}+\frac{1}{2}B_{m [\sigma} F_{\lambda\tau]}^m
\Big)
\nn \\
F_{\lambda\tau}^m&=&\partial_\lambda A_\tau^m-\partial_\tau
A_\lambda^m \\
F_{\lambda\tau m}&=&\partial_\lambda B_{m\tau}-\partial_\tau
B_{m\lambda} ~. \nn
\ea
We may now combine the above six scalars into the complex
axion/dilaton field $S$, the complex K\"ahler form field $T$ and
the complex structure field $U$ according to
\begin{eqnarray}
S&=&S_1+iS_2=a+iA^{-1}\nonumber\\
T&=&T_1+iT_2=b+iB^{-1}\nonumber\\
U&=&U_1+iU_2=c+iC^{-1} ~.
\end{eqnarray}
Define the matrices ${\cal M}_S$, ${\cal M}_T$ and ${\cal M}_U$
via
\be
{\cal M}_S=\frac{1}{S_2} \left(
\begin{array}{cc}
1 & S_1\\ S_1 & |S|^2
\end{array}
\right)\ , \label{eq:sl2mat}
\ee
with similar expressions for ${\cal M}_T$ and ${\cal M}_U$. We
also define the four $U(1)$ gauge fields $A^a$ by
\be
{A^1}_\mu=B_{4\mu}, \, {A^2}_\mu=B_{5\mu}, \, {A^3}_\mu=A_\mu^ 5,
\, {A^4}_\mu=-A_\mu^4.
\ee
and the 3-form
\be
H_{\mu\nu\rho}=3 \Big(\partial_{[\mu}B_{\nu\rho]} -\half
A_{[\mu}{}^T (\epsilon_T\otimes\epsilon_U){F}_{\nu\rho]} \Big) ~.
\ee
The action (\ref{H}) now becomes that of the $N=2$ $STU$ model
\cite{Duff:1995sm}
\begin{eqnarray}
S_{N=2}&=&\frac{1}{2\kappa_{4}^{2}}\int \d^4x\sqrt{g}A^{-1}\Bigl[
R_g + A^{-2} g^{\mu\nu}\partial_{\mu}A\partial_{\nu}A
-\frac{1}{12}g^{\mu\lambda}g^{\nu\tau}g^{\rho\sigma}
H_{\mu\nu\rho}H_{\lambda\tau\sigma}\nonumber\\
&&\kern9.3em +\frac{1}{4}\Tr(\partial{\cal M}_T{}^{-1}\partial
{\cal M}_T)
+\frac{1}{4}\Tr(\partial{\cal M}_U{}^{-1}\partial {\cal
M}_U)\nonumber\\
&&\kern9.3em -\frac{1}{4}{F_S}_{\mu\nu}{}^T({\cal M}_T \otimes
{\cal M}_U){F_S}^{\mu\nu} \Bigr]\ . \la{S}
\end{eqnarray}
The action is invariant under $\SL(2,\Rset)_{T}\times \SL(2,\Rset)_{U}$ and
the equations of motion have an additional $\SL(2,\Rset)_{S}$ .

It is now of interest to see how the original reversal
of the $D=6$ metric and 2-form
\begin{equation}
G_{MN}\rightarrow -\,G_{MN}~~~~~~B_{MN}\rightarrow -\,B_{MN}
\la{6D}
\end{equation}
manifests itself in $D=4$. The transformation property of the
$B_{MN}$-field is forced on us by supersymmetry. From (\ref{metricred}), (\ref{scalars}) and
(\ref{eq:sl2mat}) we find
\ba
g_{\mu\nu} &\longrightarrow & -g_{\mu\nu} \la{4D} \\
A^{1,2}_{\mu} &\longrightarrow & -A^{1,2}_{\mu} \\
A^{3,4}_{\mu} &\longrightarrow & A^{3,4}_{\mu} \\ S
&\longrightarrow & - S \\
T &\longrightarrow & - T \\
U &\longrightarrow & U
\ea
Thus, remarkably, the noninvariance  under
metric reversal is compensated by reversing the sign\footnote{This is
reminiscent of flipping the sign of
Newton's constant \cite{Quiros:2004ge} but here we are transforming a
field,
not a constant.} of the
shifted dilaton $A$.  In other words, the $D=4$ Einstein metric
(\ref{4DEinstein}) does not change sign. Similarly, the
non-invariance of the Yang-Mills term is compensated by reversing
the sign of $T$.

In the quantum theory $\SL(2,\Rset)$ is restricted to 
$\SL(2,\Zset)$.  It may seem unusual to find an action of a discrete 
group $\SL(2,\Zset)$ on the complex
parameters $S$ and $T$ which take values not restricted to the
upper half of the complex plane. Interestingly enough, such a situation
was also recently encountered in \cite{Kapustin:2006pk}. The context was
somewhat different but also involved orientation reversal.

\subsection{Kaluza-Klein mass terms}

Although mass terms are forbidden in the $D=4k+2$ gravitational
theories, they may nevertheless appear in lower dimensions \`a la
Kaluza-Klein. To see this consider a massless scalar whose field
equation is
signature reversal invariant:
\be
G^{MN}\nabla_{M}\nabla_{N}\phi= 0~.
\ee
Now consider a product manifold $M \times K$ and decompose the
coordinates
as $x^{M}=(x^{\mu},y^{m})$. Then
\be
(G^{\mu\nu}\nabla_{\mu}\nabla_{\nu}+G^{mn}\nabla_{m}\nabla_{n})\phi(x,y)
= 0 ~.
\ee
Fourier expanding $\phi(x,y)$ in terms of harmonics on the
internal manifold $K$, we see that the mass matrix will transform
the same way as the kinetic term under signature invariance, thus
preserving the symmetry.

\subsection{The cosmological constant}

Although a cosmological term $ \int \Lambda\sqrt{G} \d^D x $ is
forbidden in $D=4k+2$, a cosmological constant may nevertheless
arise from the vev of an antisymmetric tensor field strength
\cite{Duff:1980qv,Aurilia:1980xj}
\be
\Big\langle \sqrt{g} g^{\mu_{1}\nu_{1}}g^{\mu_{2}\nu_{2}}\ldots
g^{\mu_{n}\nu_{n}} F_{\mu_{1}\mu_{2}\ldots \mu_{n}} \Big\rangle \sim
\epsilon^{{\nu_{1}\nu_{2}\ldots \nu_{n}}} ~.
\ee
 For example, by setting
\be
F_{5} \sim \epsilon_{5}+ \star \epsilon_{5}
\ee
we could obtain $AdS^{5}\times S^{5}$ from a Freund-Rubin
\cite{Freund:1980xh} compactification of Type IIB in $D=10$.

\section{Conclusions}

Signature reversal invariance favours two kinds of theory:
Yang-Mills theories of the Euclidean type in $D=4k$, and
gravitational theories of the Minkowskian type in $D=4k+2$,
although Yang-Mills interactions may appear in the latter after
spontaneous compactification.

In many ways Type IIB supergravity is the archetypal bicoastal
theory, making use of all the ingredients: $S-T=4k'$, $T$ odd,
$D=4k+2$, no mass or cosmological terms, chiral, with gravitinos
and dilatinos of opposite chirality, involving field strengths
only of odd rank which are self-dual in the generalized sense of
section \ref{Yukawa}.  In fact it also works in the midwest and
points in between since it is invariant under metric reversal in
all its various signatures:  (9,1), (7,3), (5,5), (3,7) and (1,9).
If Type IIB supergravity did not already exist, signature reversal
invariance would have forced us to invent it. However, by virtue
of its $AdS_{5} \times S^{5}$ vacuum, it also clearly illustrates
that this symmetry does not rule out a lower-dimensional
cosmological constant. The situation with the Type IIB string is
equally interesting, with some unexpected consequences. Signature
reversal in string theory, together with reversal of the string 
coupling field, ${\Hdila}(x)$ is the subject of an accompanying paper 
\cite{Duff:2006}.

Chiral supergravity in $D=6$ (for example the $(2,0)$ theory obtained by
compactifying Type IIB on K3) is equally archetypically
bicoastal, for all the same reasons, including self-duality. The
latter property means that such theories do not possess a Lorentz
invariant action principle. It could be argued that they
occupy a privileged position in that with no action, there is
truly no way for your friends to guess what signature you had in
mind when writing down the field equations.

Although the flipping of the sign of the metric tensor may leave the
equations of motion invariant, a choice has to be made when choosing
the boundary conditions. A metric vacuum expectation value
\be
\langle G_{MN}(x) \rangle = \eta_{MN}
\ee
breaks the reversal symmetry spontaneously. Similarly
the dilaton field expectation value
\be
\langle {\Hdila}(x) \rangle = g_{s}
\ee
breaks spontaneously the sign reversal ${\Hdila} \rightarrow
-{\Hdila}$ to be discussed in the accompanying paper
\cite{Duff:2006}. In this context, it may be worth reviving
speculations about a phase of quantum gravity or string theory in
which these expectation values vanish and the symmetries are
restored, notwithstanding the noninvertibility
\cite{Deser:2006db}.

\appendix

\section{Fermion representations and Clifford algebra}
\label{LocalFermi}

Consider a metric $G_{MN}$ with $S$ positive and $T$ negative
eigenvalues. In this appendix we shall now discuss the structure
of the Clifford algebra $\Cliff(\Rset^{S,T})$.

The structure of the full Clifford algebra depends on the
signature of the metric; in general $\Cliff(\Rset^{S,T})$ differs
from $\Cliff(\Rset^{T,S})$. Only if the difference $S-T$ is a
multiple of four does it happen  that the two algebras in opposite
signatures are isomorphic. Then, denoting $D=S+T$,
\ba
\Cliff({\Rset^{S,T}}) &=& \begin{cases} \Mat_n (\Rset) ~,  n =
2^{\half D}  & \text{ for
} S-T = 8k' \\
 \Mat_n (\Hset) ~,  n = 2^{\half D -1} & \text{ for
} S-T = 8k' + 4
\end{cases} ~.
\ea
In particular, the four-dimensional Minkowski signature gives rise
to un-isomorphic Clifford algebras in the two signatures. Examples
where the Clifford algebras are the same irrespective the
signature are the $D=4k+2$ dimensional Minkowski spaces where
self-dual tensor theories have odd field strengths, and the
symmetric cases $S=T$ with an equal number of space and time
directions.

In even dimensions the chirality operator anticommutes with the generators of the
Clifford algebra. We can decompose the full algebra into two
algebras span by even, resp.~odd, products of the generators
\ba
\Cliff({\Rset^{S,T}}) &=&  \Cliff^\text{even}({\Rset^{S,T}})
\oplus \Cliff^\text{odd}({\Rset^{S,T}})  ~.
\ea
Note that the even part is isomorphic to that in the opposite
signature \ba \Cliff^\text{even}({\Rset^{S,T}})  &\simeq&
\Cliff^\text{even}({\Rset^{T,S}}) \label{spineq} \ea in any
dimensionality. The various symmetry groups are subgroups of these
algebras
\ba
\Spin(S,T) &\subset& \Cliff^{\text{even}}(\Rset^{S,T}) \\
\Pin(S,T) &\subset& \Cliff(\Rset^{S,T}) ~. \ea It follows from the
isomorphism of the algebras (\ref{spineq}) that \ba \Spin(S,T)
&\simeq &  \Spin(T,S)  ~. \ea In the special dimensions where $S-T
= 0 \mod 4$ we have also \ba \Pin(S,T) &\simeq &  \Pin(T,S) ~, \ea
which is not true in general.

\begin{table}
\begin{center}
\begin{tabular}{|c|cl|}
\hline
$T-S \mod 8$ & $N$ & $\Cliff(\Rset^{S,T}) \simeq$ \\
\hline 0,6 & $2^{\half D}$ &  $\Mat_N(\Rset)$ \\
2,4 & $2^{\half (D - 2)}$   & $\Mat_N(\Hset)$ \\
 1,5 & $2^{\half (D - 1)}$ & $\Mat_N(\Cset)$ \\
 3 & $2^{\half (D - 3) }$ & $\Mat_N(\Hset) \oplus \Mat_N(\Hset)$ \\
7 & $2^{\half (D - 1) }$ & $\Mat_N(\Rset) \oplus \Mat_N(\Rset)$ \\
\hline
\end{tabular}
\end{center}
\caption{The isomorphism classes of Clifford algebras, $D=S+T$. } 
\label{tab2}
\end{table}

\renewcommand{\arraystretch}{1.3}
\renewcommand{\tabcolsep}{6pt}
\begin{table}
\begin{center}
\begin{tabular}{||c|cc|cc|c||}
\hline \hline
& \multicolumn{2}{c|}{Euclidean}  & \multicolumn{2}{c|}{Minkowskian}
& Complexified \\
\hline
\multirow{2}{16pt}{$D$}  & Space-like & Time-like    & Space-like &
Time-like   & Complexified \\
& $(D,0)$ & $(0,D)$ & $(D-1,1)$ & $(1,D-1)$ & $D$  \\
\hline 1 & $\Rset \oplus \Rset $ & $\Cset$ & $\Cset$ &
$\Rset \oplus \Rset $  & $\Cset \oplus \Cset$ \\
2 & $\Rset(2)$ & $\Hset$  &  $\Rset(2)$ & $\Rset(2)$  & $\Cset(2)$  \\
3 & $\Cset(2)$ & $\Hset \oplus \Hset$ &   $\Rset(2) \oplus \Rset(2) $
& $\Cset(2)$ & $\Cset(2) \oplus \Cset(2)$  \\
4 & $\Hset(2)$ & $\Hset(2)$ &  $\Rset(4)$ & $\Hset(2)$ & $\Cset(4)$
\\
5 &  $\Hset(2) \oplus \Hset(2) $ &  $\Cset(4)$ & $\Cset(4)$ &
$\Hset(2) \oplus \Hset(2) $  & $\Cset(4)\oplus \Cset(4)$  \\
6 & $\Hset(4)$ & $\Rset(8)$ &  $\Hset(4)$ & $\Hset(4)$& $\Cset(8)$ \\
7 & $\Cset(8)$ &  $\Rset(8) \oplus \Rset(8) $ & $\Hset(4) \oplus
\Hset(4) $  & $\Hset(8)$ & $\Cset(8)\oplus \Cset(8)$ \\
8 & $\Rset(16)$ &$\Rset(16)$& $\Cset(8)$ & $\Rset(16)$ &
$\Cset(16)$ \\
9 & $\Rset(16) \oplus \Rset(16) $ & $\Cset(16)$ & $\Cset(16)$ &
$\Rset(16) \oplus \Rset(16) $  & $\Cset(16) \oplus \Cset(16)$\\
10 & $\Rset(32)$ & $\Hset(16)$  &  $\Rset(32)$ & $\Rset(32)$  &
$\Cset(32)$\\
11 & $\Cset(32)$ & $\Hset(16) \oplus \Hset(16)$ &   $\Rset(32) \oplus
\Rset(32) $ & $\Cset(32)$ & $\Cset(32) \oplus \Cset(32)$\\
12 & $\Hset(32)$ & $\Hset(32)$ &  $\Rset(64)$ & $\Hset(32)$ &
$\Cset(64)$\\
\hline \hline
\end{tabular}
\end{center}
\caption{Clifford algebras for Minkowskian and Euclidean
signatures, where $\Rset(n) \equiv \Mat_n(\Rset)$ \etc. }
\la{tab3}
\end{table}

\newcommand{\tinyA}{\scriptsize}
\newcommand{\tinyB}{\scriptsize}
\begin{landscape}
\renewcommand{\arraystretch}{1.7}
\renewcommand{\tabcolsep}{0pt}
\begin{table}
\begin{center}
\begin{tabular}{||cc|c|cccccccccccc||}
\hline \hline
\multicolumn{3}{||c|}{} & \multicolumn{12}{c||}{Space-like dimensions} \\
\cline{4-15}
\multicolumn{3}{||c|}{} & $0$ & $1$ & $2$ & $3$  & $4$ & $5$ & $6$ & $7$ & $8$ & $9$ & $10$ & $11$ \\
\hline \multirow{9}{18pt}{~\begin{sideways}\mbox{Time-like
dimensions}\end{sideways}}
&& \ 11 \ &   \tinyB  \; $\Hset(16) \oplus \Hset(16)$ &&& &&&&& &&& \\
&& 10 & \tinyA  $\Hset(16)$ & \tinyA  $\Cset(32)$ && &&&&& &&& \\
&& 9 & \tinyA  $\Cset(16)$ & \tinyA  \fbox{ $\Rset(32)$} &  \tinyB  {$\Rset(32) \oplus \Rset(32)$} & &&&&& &&& \\
&& 8 & \tinyA  \fbox{ $\Rset(16)$ } &  \tinyB   $\Rset(16) \oplus \Rset(16)$ & \tinyA  $\Rset(32)$ & \tinyA  $\Cset(32)$ &&&&& &&& \\
&&7& \tinyB   $\Rset(8) \oplus \Rset(8)$ &  \tinyA  $\Rset(16)$ & \tinyA  $\Cset(16)$  & \tinyA  \fbox{$\Hset(16)$ }&   \tinyB   $\Hset(16) \oplus \Hset(16)$ &&&&  &&&  \\
&&6& \tinyA  $\Rset(8)$ & \tinyA  $\Cset(8)$  & \tinyA  \fbox{$\Hset(8)$} &  \tinyB   $\Hset(8) \oplus \Hset(8)$ & \tinyA  $\Hset(16)$ & \tinyA  $\Cset(32)$ &&& &&& \\
&&5& \tinyA  $\Cset(4)$  & \tinyA  \fbox{$\Hset(4)$} & \tinyB
$\Hset(4) \oplus \Hset(4)$ & \tinyA  $\Hset(8)$ & \tinyA
$\Cset(16)$ & \tinyA  \fbox{$\Rset(32)$} &  \tinyB   $\Rset(32)
\oplus \Rset(32)$  &&  &&&
\\
&&4& \tinyA  \fbox{$\Hset(2)$} & \tinyB   $ \tinyB  \Hset(2) \oplus \Hset(2)$ & \tinyA  $\Hset(4)$ & \tinyA  $\Cset(8)$ & \tinyA  \fbox{$\Rset(16)$ }& \tinyB   $\Rset(16) \oplus \Rset(16)$  & \tinyA  $\Rset(32)$ & \tinyA  $\Cset(32)$ & &&& \\
&&3&  \tinyB   $\Hset \oplus \Hset$ & \tinyA  $\Hset(2)$ & \tinyA  $\Cset(4)$ & \tinyA  \fbox{$\Rset(8)$ }&  \tinyB   $\Rset(8) \oplus \Rset(8)$  & \tinyA  $\Rset(16)$ & \tinyA  $\Cset(16)$ & \tinyA   \fbox{$\Hset(16)$} & \tinyB  $ \Hset(16) \oplus \Hset(16)$ &&&   \\
&&2& \tinyA   $\Hset$ & \tinyA  $\Cset(2)$ & \tinyA   \fbox{$\Rset(4)$} &  \tinyB   $ \Rset(4) \oplus \Rset(4)$  & \tinyA  $\Rset(8)$ & \tinyA  $\Cset(8)$ & \tinyA    \fbox{ $\Hset(8)$ } &   \tinyB  $\Hset(8) \oplus \Hset(8)$ & \tinyA    {$\Hset(16)$}   & \tinyA  $\Cset(32)$  &&  \\
&&1& \tinyA  $\Cset$ & \tinyA   \fbox{$\Rset(2)$ } &   \tinyB  $\Rset(2) \oplus \Rset(2)$  & \tinyA  $\Rset(4)$ & \tinyA  $\Cset(4)$ & \tinyA   \fbox{$\Hset(4)$} &  \tinyB   $\Hset(4) \oplus \Hset(4)$ & \tinyA $\Hset(8)$ & \tinyA  $\Cset(16)$  & \tinyA  \fbox{ $\Rset(32)$}  & \tinyB  $\Rset(32) \oplus \Rset(32)$ & \\
&&0& \tinyA 0 &  \tinyB  $\Rset \oplus \Rset$  & \tinyA  $\Rset(2)$ & \tinyA  $\Cset(2)$ & \tinyA  \fbox{ $\Hset(2)$ }&   \tinyB  $\Hset(2) \oplus \Hset(2)$ & \tinyA $\Hset(4)$ & \tinyA  $\Cset(8)$ & \tinyA   \fbox{ $\Rset(16)$}  &  \tinyB  $\Rset(16) \oplus \Rset(16)$   & \tinyA  $\Rset(32)$ & \tinyA  $\Cset(32)$ \;  \\
\hline \hline
\end{tabular}
\end{center}
\caption{Isomorphism classes of Clifford algebras $D < 12$ in
arbitrary signatures, where $\Rset(n) \equiv \Mat_n(\Rset)$ \etc.
Boxed entries are symmetric in signature reversal. }
\la{tab4}
\end{table}
\end{landscape}

Fermions belong to representations of the Clifford algebra. Let us
concentrate in $S-T =4k'$, so that the Clifford algebra can be
thought of as an algebra of real (or quaternionic) matrices of a
given dimensionality. There the representations of the fermions
that do not satisfy a chirality condition are referred to as
Majorana (resp.~quaternionic Majorana) fermions. In Mathematics
Literature these representations are usually referred to as {\em
pinors} $P$.

In these dimensionalities  $S-T =4k'$,  the chirality operator is
nilpotent $\Gamma^2=\unit$, and the pinor representations $P$
split to positive and negative eigenspaces $S_\pm$ with respect to
it $P=S_+ \oplus S_-$. These (quaternionic) Majorana-Weyl fermions
are in Mathematics Literature usually referred to as {\em
spinors}. The even part of the Clifford algebra acts diagonally on
$S_+ \oplus S_-$, whereas the odd part does not. Note that a
similar split does not occur in other even dimensionalities, as
there the two chiralities are typically related by complex
conjugation.

\section{Real structure}
\la{real}

In most literature the gamma matrices $\Gamma_M(x)
\equiv {e}_M^{~~A}(x)\Gamma_A$ are defined in terms of a set of
constant reference matrices $\Gamma_A$ that satisfy $ \{ \Gamma_A,
~ \Gamma_B \} = 2 \eta_{AB}$. The defining equation for  vielbeins
${e}_M^{~~A} $ is
\ba
{e}_M^{~~A} ~ {e}_N^{~~B} ~ \eta_{AB} &=& G_{MN} ~.
\ea

One may ask whether one could account for the change of signature
$G_{MN} \longrightarrow - G_{MN}$ by a redefinition of vielbeins
\ba
\bar{e}_M^{~~A} &\equiv & {e}_M^{~~B} ~ J_B^{~A} ~.
\ea
This requires
\ba
J_C^{~A}   J_D^{~B} ~ \eta_{AB} &=& - \eta_{CD} \\
J_A^{~C}   J_C^{~B} &=&  + \delta_A^{~B}  ~. \label{in}
\ea
When indices are lowered with $\eta_{AB}$, this means that
$J_{AB}$ is antisymmetric, and $\det J = \pm 1$.

One can show that neither in Euclidean, nor in $D>2$ Minkowski signature is there a non-trivial solution to these equations. The problem really is the involution (\ref{in}). To see this, let us write
\ba
(J_{AB}) &=& \left(
\begin{array}{cc}
A & B \\
-B^{T} & C
\end{array}
\right) \\
(J_{AB}\eta^{BC}) &=& \left(
\begin{array}{cc}
-A & B \\
B^{T} & C
\end{array}
\right)
~,
\ea
where $A$ and $C$ are antisymmetric $T \times T$ resp.~$S \times S$~matrices. The square of this is
\ba
\unit_{T} \oplus \unit_{S} &=&
J \eta^{-1}J \eta^{-1} \\
&=&
\left(
\begin{array}{cc}
A^{2} + BB^{T} & -AB+BC \\
-B^{T}A+CB^{T} & B^{T}B+C^{2}
\end{array}
\right) ~.
\ea

Let us consider the different cases for $T=0,1$ and $T>1$ respectively.
\begin{itemize}
\item In Euclidean signature $T=0$, so that $A=B=0$ and
\ba
\unit_{S} &=& C^{2} = - C^{T}C \leq 0 ~.
\ea
This implies
\ba
{S} &=& - \tr C^{T}C \leq 0 ~;
\ea
this is a contradiction, and it follows that there are no matrices $C$ that satisfy the constraints imposed   above.
\item In Minkowskian signature $T=1$ and $A$ is an antisymmetric real number, hence zero $A=0$. We have
\ba
1 \oplus \unit_{S} &=& \left(
\begin{array}{cc}
BB^{T} & BC \\
CB^{T} & B^{T}B+C^{2}
\end{array}
\right) \\ ~.
\ea
It follows
\ba
1 &=& BB^{T} \\
S &=& \tr (B^{T}B+C^{2}) = BB^{T} - \tr C^{T}C
\ea
so that
\ba
S-1 &=& -\tr C^{T}C \leq 0 ~.
\ea
This means that $S \leq 1$, and there is a real structure available in a Minkowski space only for $(S,T) =(1,1)$.
\item
This argument does not generalise to higher $T>1$ as then the equation
\ba
\unit_{T} &=& -A^{T}A + BB^{T}
\ea
has more than one solution.
\end{itemize}

More generally, however, in $\Rset^{{p,p}}$ we can choose
\ba
(\eta_{AB}) &=&  \sigma_{3} \otimes \unit_{p}\\
(J_A^{~B})  &=& \sigma_{1} \otimes \unit_{p} \\
(J_{AB})   &=&  i\sigma_{2} \otimes \unit_{p} ~,
\ea
which solves the two constraints.



\newpage


\begin{thebibliography}{99}


\bibitem{Erdem:2004yd}
  R.~Erdem,
  ``A symmetry for vanishing cosmological constant in an extra
dimensional toy model,''
  Phys.\ Lett.\ B {\bf 621}, 11 (2005)
  [arXiv:hep-th/0410063].

\bibitem{Erdem:2006qk}
  R.~Erdem,
  ``A symmetry for vanishing cosmological constant: Another
realization,''
  [arXiv:gr-qc/0603080].

\bibitem{Quiros:2004ge}
  I.~Quiros,
  ``Symmetry relating gravity with antigravity: A possible resolution
of  the cosmological constant problem?''
  [arXiv:gr-qc/0411064].

\bibitem{Nobbenhuis:2004wn}
  S.~Nobbenhuis,
  ``Categorizing different approaches to the cosmological constant
problem,''
  [arXiv:gr-qc/0411093].

\bibitem{'tHooft:2006rs}
  G.~'t Hooft and S.~Nobbenhuis,
  ``Invariance under complex transformations, and its relevance to the
  cosmological constant problem,''
  [arXiv:gr-qc/0602076].

\bibitem{Kaplan:2005rr}
  D.~E.~Kaplan and R.~Sundrum,
  ``A symmetry for the cosmological constant,''
  [arXiv:hep-th/0505265].

\bibitem{Bonelli:2000tz}
  G.~Bonelli and A.M.~Boyarsky,
   ``Six dimensional topological gravity and the cosmological constant
   problem,''
  Phys.\ Lett.\ B {\bf 490} (2000) 147
  [arXiv:hep-th/0004058].

\bibitem{Carlip:1988gw}
  S.~Carlip and C.~DeWitt-Morette,
  ``Where the sign of the metric makes a difference,''
  Phys.\ Rev.\ Lett.\  {\bf 60} (1988) 1599.

\bibitem{Berg:2000ne}
  M.~Berg, C.~DeWitt-Morette, S.~Gwo and E.~Kramer,
  ``The Pin groups in physics: C, P, and T,''
  Rev.\ Math.\ Phys.\  {\bf 13}, 953 (2001)
  [arXiv:math-ph/0012006].

\bibitem{Linde:1988ws}
  A.D.~Linde,
  ``The universe multiplication and the cosmological constant
problem,''
  Phys.\ Lett.\ B {\bf 200}, 272 (1988).

\bibitem{Duff:2006}
  M.J.~Duff and J.~Kalkkinen,
  ``Signature and coupling reversal in string theory,''
  [arXiv:hep-th/0605274].

\bibitem{RH} F.~Reese Harvey, ``Spinors and calibrations,''
(Academic Press , 1990)

\bibitem{Bergshoeff:1981um}
  E.~Bergshoeff, M.~de Roo, B.~de Wit and P.~van Nieuwenhuizen,
  ``Ten-Dimensional Maxwell-Einstein supergravity, its currents, and the issue of its auxiliary fields,''
  Nucl.\ Phys.\ B {\bf 195} (1982) 97.
  
\bibitem{Polchinski:1998rr}
  J.~Polchinski,
  ``String theory. Vol. 2: Superstring theory and beyond.''

\bibitem{Schwarz:1983qr}
  J.H.~Schwarz,
  ``Covariant field equations of chiral $N=2$ $D=10$ supergravity,''
  Nucl.\ Phys.\ B {\bf 226}, 269 (1983).


\bibitem{Duff:1995sm}
  M.J.~Duff, J.T.~Liu and J.~Rahmfeld,
  ``Four-dimensional string/string/string triality,''
  Nucl.\ Phys.\ B {\bf 459} (1996) 125
  [arXiv:hep-th/9508094].

\bibitem{Kapustin:2006pk}
  A.~Kapustin and E.~Witten,
  ``Electric-magnetic duality and the geometric Langlands program,''
  [arXiv:hep-th/0604151].
  
\bibitem{Duff:1980qv}
  M.J.~Duff and P.~van Nieuwenhuizen,
  ``Quantum inequivalence of different field representations,''
  Phys.\ Lett.\ B {\bf 94}, 179 (1980).


\bibitem{Aurilia:1980xj}
  A.~Aurilia, H.~Nicolai and P.K.~Townsend,
  ``Hidden constants: The theta parameter of QCD and the cosmological constant of $N=8$ supergravity,''
  Nucl.\ Phys.\ B {\bf 176}, 509 (1980).

\bibitem{Freund:1980xh}
  P.G.O.~Freund and M.A.~Rubin,
  ``Dynamics of dimensional reduction,''
  Phys.\ Lett.\ B {\bf 97}, 233 (1980).

\bibitem{Deser:2006db}
  S.~Deser,
  ``Why is the metric invertible?''
  [arXiv:gr-qc/0603125].


\end{thebibliography}
\end{document}